\documentclass{PoS}

\usepackage{dcolumn}
\usepackage{bm}
\usepackage{graphicx}
\usepackage{epstopdf}
\usepackage{epsfig}
\usepackage{color}
\usepackage{wasysym}
\usepackage{twoopt}
\usepackage{verbatim}
\usepackage{dashrule}

\def\Journal#1#2#3#4{{#4}, {#1}, {#2}, #3} 

\newcommand{\GALPROP}{\textsf{GALPROP}}
\newcommand{\WNEW}{{\textsf{WNEW}}}
\newcommand{\YIELDX}{\textsf{YIELDX}}
\newcommand{\XS}{\textsf{XS}}
\newcommand{\XSs}{\textsf{XS's}}
\newcommand{\PF}{\textsf{P}$\rightarrow$\textsf{F}}
\newcommand{\Hyd}{\textsf{H}}
\newcommand{\He}{\textsf{He}}

\newcommand{\Be}{\textsf{Be}}
\newcommand{\B}{\textsf{B}}
\newcommand{\C}{\textsf{C}}
\newcommand{\N}{\textsf{N}}
\newcommand{\Oxy}{\textsf{O}}
\newcommand{\Ne}{\textsf{Ne}}
\newcommand{\Mg}{\textsf{Mg}}

\newcommand{\Si}{\textsf{Si}}
\newcommand{\Fe}{\textsf{Fe}}
\newcommand{\BeB}{\textsf{Be}/\textsf{B}}  
\newcommand{\BC}{\textsf{B}/\textsf{C}}

\newcommand{\etal}{et alii}

\newcommand{\AMS}{\textsf{AMS}}
\newcommand{\eg}{\textit{e.g.}} 
\newcommand{\ie}{\textit{i.e.}}

\def\Journal#1#2#3#4{{#4}, {#1}, {#2}, #3}

\def\citep#1{{\cite{#1}}}
\def\citet#1{{\cite{#1}}}

\title{Fragmentation cross-sections and model uncertainties in Cosmic Ray propagation physics}
\ShortTitle{Wanted! Nuclear data for CR Physics with AMS}

\author{\speaker{Nicola Tomassetti}\\ 
LPSC, Universit\'e Grenoble-Alpes, CNRS/IN2P3, F-38026 Grenoble, France; email: nicola.tomassetti@lpsc.in2p3.ch\\
        E-mail: \email{nicola.tomassetti@lpsc.in2p3.fr}}

\abstract{Abundances and energy spectra of cosmic ray nuclei are being measured with high accuracy by the AMS experiment. These observations can provide tight constraints to the propagation models of galactic cosmic rays. In the view of the release of these data, I present an evaluation of the model uncertainties associated to the cross-sections for secondary production of Li-Be-B nuclei in cosmic rays. I discuss the role of cross section uncertainties in the calculation of the boron-to-carbon and beryllium-to-boron ratios, as well as their impact in the determination of the cosmic-ray transport parameters.}

\FullConference{The 34th International Cosmic Ray Conference,\\
		30 July- 6 August, 2015\\
		The Hague, The Netherlands}

\begin{document}

\section{Introduction} 
%
The determination of the Cosmic Ray (CR) transport parameters is a central question to astrophysics.
Models of CR propagation accounts for particle diffusion off magnetic 
turbulence and interactions with the interstellar medium (ISM). 
\textit{Primary} CRs such as \C-\N-\Oxy{} nuclei are those accelerated in supernova remnants.
\textit{Secondary} nuclei are created by collisions (or decays) of primary CRs off the ISM gas.  
\textit{Secondary-to-primary} ratios of stable nuclei, and notably the \BC{} ratio,
are used to determine the diffusion coefficient, $D$, and the half-size propagation region, $L$ \citep{Strong2007,Maurin2001}.
The diffusion coefficient is usually expressed as $D \propto D_{0}R^{\delta}$. 
The \BC{} ratio is sensitive to $\delta$ and to the $D_{0}/L$ ratio.
The degeneracy between $D_{0}$ and $L$ can be resolved using the data on \textit{unstable-to-stable} 
isotopic ratios, \eg, the $^{10}$\Be/$^{9}$\Be{} ratio. 
The \textit{decaying-to-decayed} elemental ratio \BeB{} at $\sim$\,1--10\,GeV 
can also be used in place of the isotopic ratio $^{10}$\Be/$^{9}$\Be{} \citep{WebberSoutoul1998}. 
Thus, the data combination \BC{}\,$+$\,\BeB{} may allow to extract the basic information on the CR transport.
An application on this study is the search of dark matter annihilation signals, 
for which resolving the $L/D_{0}$ degeneracy is of great importance.
The spectra of \B{} and \Be{} nuclei are now being precisely measured by the Alpha Magnetic Spectrometer 
(\AMS) on the \textit{International Space Station} (ISS).
Recent results on CR protons \citep{Aguilar2015} and preliminary light nuclei data 
at GeV -- TeV energies \citep{AMSDays2015} have been recently presented.
With the \AMS{} standards of precision, it is timely to review the major uncertainties of the model predictions.
In particular, the parameters extraction relies on the secondary production calculations for \Be{} and \B{} nuclei, 
which depend on several cross-section (\XS) estimates.
Our understanding of the fragmentation \XSs{} relies on the available nuclear data.
Thus, the accuracy of the inferred transport parameters is directly linked to the quality 
of the fragmentation \XS{} measurements.
In this paper, I estimate the impact of the nuclear uncertainties in the CR parameter extraction
within the precision that we expect from \AMS. In particular, my study is focused on simulated 
data on the ratios \BC{} and \BeB, and their connection with the $D_{0}/L$ degeneracy.
For this purpose, a survey of the literature was done in order to collect several \Be{} and \B{} 
production \XSs{} from \B-\C-\N-\Oxy{} collisions off hydrogen.
These data are used to constrain the \XS{} formulae in order to obtain an estimate of their uncertainties.   
The resulting \XS{} uncertainties are then converted into model uncertainties of the predicted ratios, and
eventually into the uncertainties on the relevant parameters that can be inferred by \AMS.

\section{Cosmic-Ray Transport and Interactions in the Galaxy}  
%
In this work I employ the \textit{diffusive-reacceleration} model implemented under the numerical code \GALPROP{},
which solves the CR propagation equation for a given set of input parameters \citep{Strong2007}.
The transport equation for a CR species $j$ is expressed as: 
\begin{equation}\label{Eq::DiffusionTransport}
  \partial_{t} {\psi}_{j} = 
  q_{j}+
  \vec{\nabla}\cdot\left[D\vec{\nabla}{\psi}_j\right] - 
      {{\psi}_j}{\Gamma_{j}} +
      \partial_{p}\left[ p^{2} D_{pp} \partial_{p}{p^{-2}} - \dot{p}_{j} \right] {\psi}_j  
\end{equation}
where $\psi_{j}=\frac{dN_{j}}{dV dp}$ is the particle density per unit of momentum $p$. 
The CR acceleration in primary sources is described as $q_{j}^{\rm pri}\propto R^{-\nu}$,
while the secondary production term is $q_{j}^{\rm sec} = \sum_{k} {\psi}_{k} \Gamma_{k\rightarrow j}$,
for fragmentation/decay of $k$-type nuclei into $j$-type nuclei. 
The secondary production rate is:
\begin{equation}\label{Eq::ProductionRate}
  \Gamma_{k\rightarrow j} =  \beta_{k}c \sum_{i} n_{i} \sigma_{k\rightarrow j}^{i}(E) dE\,,
\end{equation} 
where $n_{i}$ is the number densities of the ISM nuclei, $n_{\rm H}\cong$\,0.9\,cm$^{-3}$ and $n_{\rm He}\cong$\,0.1\,cm$^{-3}$,  
and $\sigma_{k \rightarrow j}^{i}$ is the $j$-th nucleus production \XS{} at energy $E$ from $k$-nuclei destruction off the $i$-th target.
The term $\Gamma_{j}$ is the destruction rate for a cross section $\sigma_{j}^{\rm tot}$ or particle decay with
lifetime $\tau_{j}$. The diffusion coefficient $D$ is taken as $D(R)=\beta D_{0}\left(R/R_{0}\right)^{\delta}$, 
where $D_{0}$ gives its normalization at $R=R_{0}\equiv$\,4\,GV, and $\delta$ gives its rigidity dependence. 
The reacceleration is described as diffusion process acting in momentum space. 
The momentum diffusion coefficient is $D_{pp}\propto p^{2}v_{A}^{2}/D$,
where $v_{A}$ is the Alfv\'en speed of magnetic plasma waves in the ISM. 
The term $\dot{p}_{j}=dp_{j}/dt$ is the momentum loss rate for Coulomb and ionization losses.
The steady-state equation $\partial {\psi}_{j} / \partial t = 0$ is solved into a cylindrical halo of 
half-thickness $L$ with the zero-flux condition at the boundaries. 
The local interstellar spectrum for each species as function of kinetic energy per nucleon
is given by $\Phi^{\rm IS}_{j}(E) = \frac{c A}{4\pi}{\psi}_{j}(r_{\odot},p)$.
To describe the solar modulation effect, I will adopt the so-called \textit{force-field approximation} \citep{Gleeson1968}. 
After propagation, the primary primary nuclei spectra are of the type $\mathcal{P} \propto\,(L/K_{0})E^{-\nu-\delta}$
\ie, degenerated between source and transport parameters. 
The use of the \BC{} ratio allows to determine the parameter $\delta$.%
The remaining  $D_{0}$--$L$ degeneracy can be lifted using unstable isotopes such as $^{10}$\Be{} (lifetime $\tau \approx 1.5\,$Myr),
because its mean propagation length is $\lambda_{u} = \sqrt{D \gamma \tau} \ll L$ below a few GeV/nucleon.
In principle also data on the ratio \BeB{} can be used, because
it maximizes the effect of the of radioactive decay $^{10}$\Be$\rightarrow$$^{10}$\B{}+\textsf{e}$^{-}$+$\bar{\nu}_{e}$.
The \BC{} and \BeB{} ratios are currently being measured by \AMS{}.

\section{Fragmentation Cross Sections and Uncertainty Estimates} 
%
Several fragmentation \XSs{} are needed to compute the secondary production rate, 
because \Be{} or \B{} nuclei are produced by several 
\textit{projectile}$\rightarrow$\textit{fragment} combinations (\PF).
\begin{figure*}[!t]
  \includegraphics[width=0.94\textwidth]{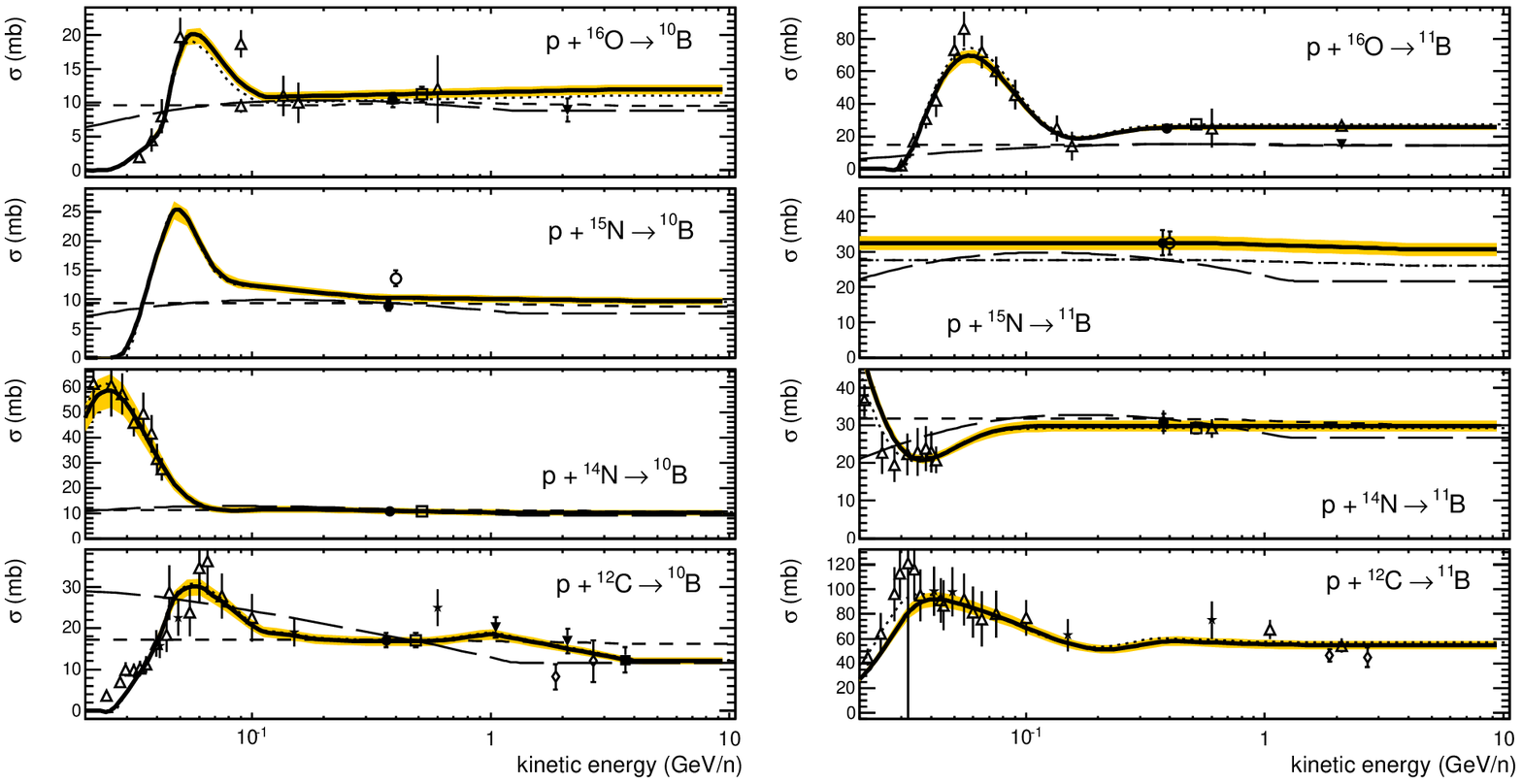} \label{Fig::ccCSBoronProd}
  \vspace{5.00mm}
  \includegraphics[width=0.94\textwidth]{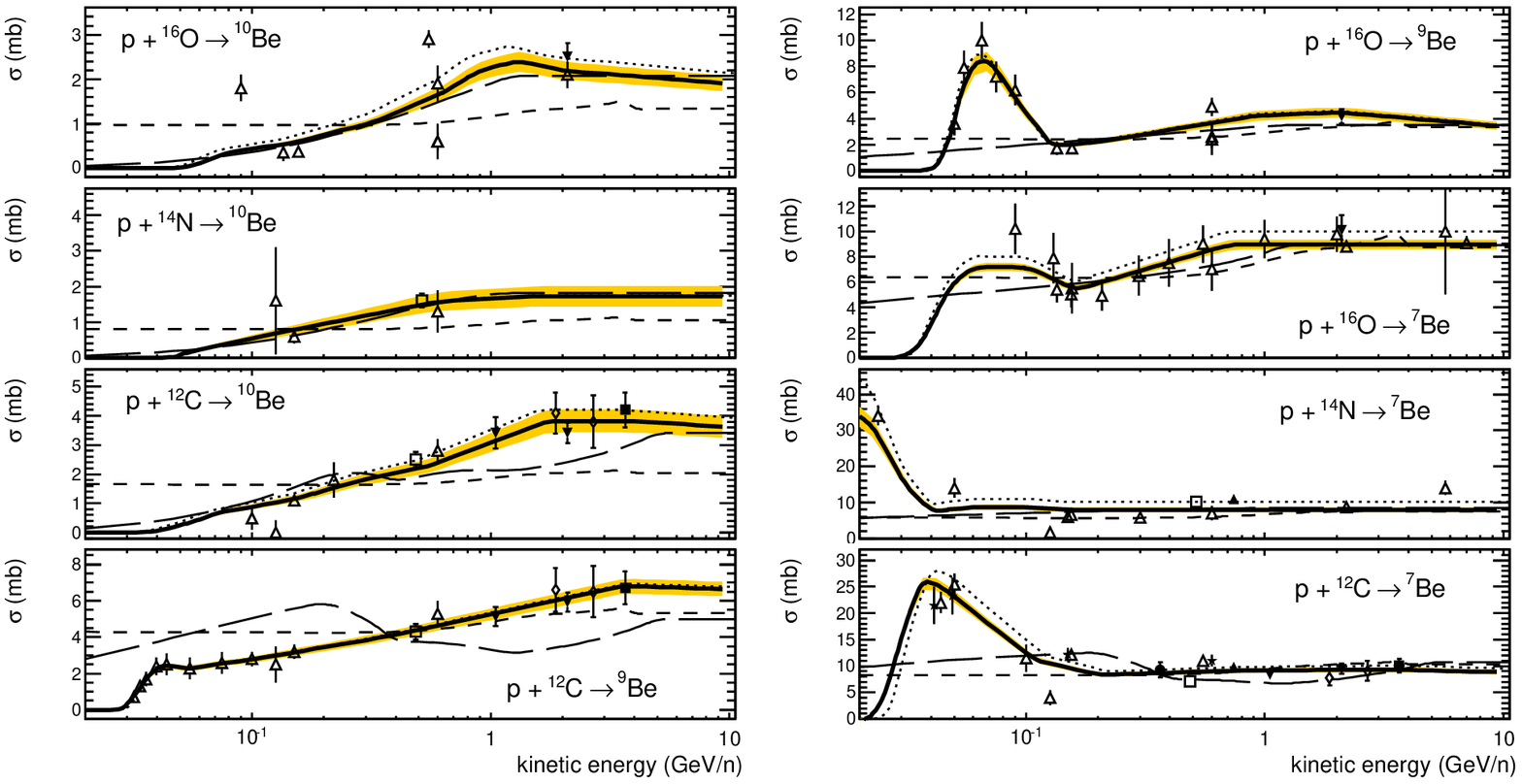}
  \caption{\footnotesize%
    Fragmentation \XSs{} for $^{10}$\B, $^{11}$\B, $^{7}$\Be{}, $^{9}$\Be, and $^{10}$\Be{} production from 
    \C-\N-\Oxy{} collisions off hydrogen. The data are from
    \citet{ReadViola1984,Webber1998,Webber1990c,Olson1983,Fontes1977,Korejwo2000,Korejwo2001,Radin1979,Ramaty1997,Webber1998prc,Raisbeck1971}.
    The lines are from the \WNEW{} (short-dashed), \YIELDX{} (long-dashed), \GALPROP{} (dotted), 
    and the re-normalized \XSs{} of this work (thick solid lines) with their uncertainty band.
  }\label{Fig::ccCSBoronProd}
\end{figure*}
Popular algorithms are \YIELDX{} \citep{Silberberg1998} 
or \WNEW{} \citep{Webber1990b,Webber1990d,Webber2003}, that provide energy-dependent 
\XSs{} off hydrogen target for several \PF{} reactions.
Under \GALPROP, the production \XSs{} come from the \texttt{CEM2k} and \texttt{LAQGSM} codes,
normalized to the data \citep{Gudima1983,Mashnik2000,Gudima2001,MoskalenkoMashnik2003}.
The \XSs{} for isotopically separated fragment/target have been measured by several experiments,
though the data are available in only narrow energy ranges.
Figure\,\ref{Fig::ccCSBoronProd} shows the data for \Be{} and \B{} isotopes from 
fragmentation of \C-\N-\Oxy{} nuclei off hydrogen at 30\,MeV/n -- 10\,GeV/n.
Beryllium is also produced via \textit{tertiary} reactions, such as \B$\rightarrow$\Be{},
that are considered in this study.
Spallation of heavier nuclei such as \Ne-\Mg-\Si{} or \Fe{} gives a minor contribution and it is not considered here.
Many reactions have an energy dependence at $E\lesssim$\,0.5\,GeV/n which is often ignored 
in CR propagation, but may be important in the context of reacceleration models (considered here) 
At energy above than a few GeV/nucleon, all the \XSs{} are nearly constant in energy. 
The data in Fig.\,\ref{Fig::ccCSBoronProd} are compared with the \XS{} formulae from \WNEW, \YIELDX, and  \GALPROP.  
Despite large discrepancies among the various formulae, the \GALPROP{} \XSs{} describe well the data.
%
%
In order to determine the \XS{} uncertainties using the data, 
I have performed a \emph{re-normalization} of the \GALPROP{} parameterizations $\sigma_{\rm G}(E)$ to the data.
For each \PF{} channel, the \XS{} has been re-fit as 
$\sigma_{\rm H}(E) = a \sigma_{\rm G}(b E)$, 
where the parameters $a$ and $b$ represent the normalization and the energy scale. 
%
%
Some of the re-evaluated \XSs{} are shown as solid lines in Fig.~\ref{Fig::ccCSBoronProd}.
The shaded bands are the estimated uncertainties.
These re-fitted \XSs{} are often close to their original values,
%
though the \Be{} production under \GALPROP{} is found to be over-estimated by a few percent. 
A \Be{} overproduction was also reported in \citet{AMS01Nuclei2010}, and it was ascribed to the production \XSs.
To account for a 10\% fraction of interstellar helium, it was applied the rescaling factor $F_{\alpha/p}$ from \citet{Ferrando1988},
and the \He-target \XSs{} are assumed to have the same relative uncertainties of the \Hyd-target \XSs.
For the total destruction I have employed the formula of \citet{Tripathi1999}.
These reactions are known with better precision and their uncertainty has a negligible impact on the \Be-\B{} propagation.
\begin{figure*}[!t] 
\includegraphics[width=0.94\textwidth]{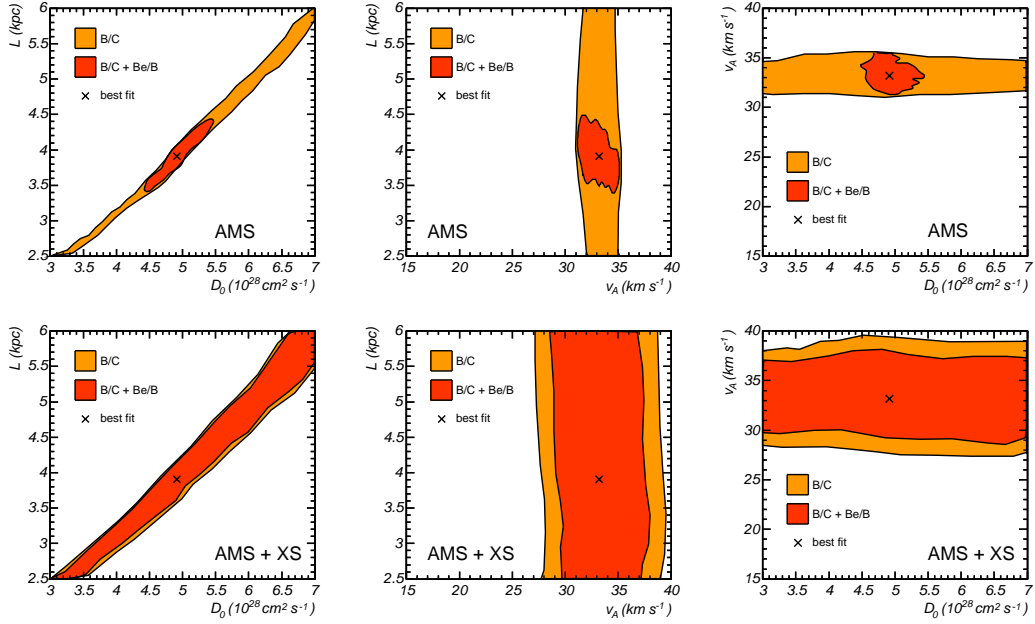}
\caption{\footnotesize%
  Top: estimation of the \AMS{} capabilities in constraining the parameters 
  $D_{0}$, $L$, and $v_{A}$ with with the \BC{} and \BeB{} ratios.
  Bottom: same as above after accounting for \emph{nuclear uncertainties} in the \Be-\B{} production rates.
}\label{Fig::ccChiSquaresAMS02}
\end{figure*}

\section{\AMS{} Physics Potential and Impact of Nuclear Uncertainties}  

The anticipated \AMS{} data for the ratios \BC{} and \BeB{} have been computed 
as in \cite{TomassettiDonato2012} using the input fluxes for $\Phi_{\rm Be}$, $\Phi_{\rm B}$, 
and $\Phi_{\rm C}$ generated with a \textit{reference model}.
In the reference model, the primary CRs injection spectra 
are taken as power-law with index $\nu=2.38$.
The diffusion parameters are $D_{0}=5 \cdot\,10^{28}$\,cm$^{2}$\,s$^{-1}$, $\delta=$\,0.38, and $v_{A}=$\,33\,km\,s$^{-1}$. 
The propagation region has half-height $L=$\,3.9\,kpc. 
The modulation parameter is taken as $\phi=550$\,MV.
For the \Be-\B{} production I have used the \XSs{} re-determined in this work, and   
their estimated uncertainties have been translated into uncertainties for the propagated fluxes and ratios. 
Typical uncertainties are $\sim$\,5\,\% for \B{} production and $\sim$\,7\--10\% for \Be{} production, 
with $\sim$\,10\% for $^{10}$\Be{} productions. 
The \Be-\B{} elements are being measured by \AMS{} at energies from $\sim$\,0.5\,GeV to $\sim$\,1\,TeV per nucleon. 
In this work, I consider the \BC{} ratio between 2 and 200\,GeV/n and the \BeB{} ratio between 1 and 100\,GeV/n. 
In this energy range, the influence of solar modulation is below $\sim$\,1\,\%. 
The expected \AMS{} ability in constraining the model parameters is first estimated
\emph{without} accounting for the nuclear uncertainties.
A grid scan is performed in the parameter space $D_{0} \times L \times v_{A}$ ,
running \GALPROP{}  3,420 times, over a $19 \times 15 \times 12$ grid. 
Thus, the resulting spectra are tri-linearly interpolated to a
finer parameter grid corresponding to 187,245 models.
From the \BC{} ratio predicted by each model, (\BC)$_{\rm mod}$, the $\chi^{2}$ is computed for the \AMS{} 
mock data, (\BC)$_{i}$, that are generated with the \emph{reference model}:
\begin{equation}\label{Eq::ChiSquare}
\chi_{\BC}^{2} = \sum_{\rm i} \left[ \frac{  (\BC)_{\rm mod} - (\BC)_{i}  }{\delta(\BC)_{i}} \right]^{2} \,
\end{equation}
Similarly, the $\chi^{2}$ is also computed for the \BeB{} ratio and for both ratios combined.
In Fig.~\ref{Fig::ccChiSquaresAMS02} (top panels) the one-sigma contour regions are shown as 
2D projections of the parameter space using the $\chi^{2}$ for the \BC{} ratio and
for the \BC{}$+$\BeB{} combined ratios.
The best-fit model is marked as ``$\times$'' on each plot. It always recovers the true reference model. 
The complementarity of the two ratios is apparent in resolving the $L$-$D_{0}$ degeneracy.
While the \BC{} ratio constrains $L$ and $D_{0}$ into a tight region of the ($L,D_{0}$) plane, 
only the combined \BC{}+\BeB{} ratios allow to determine their single values.
On the other hand, Alfv\'en speed $v_{A}$ can be determined by means of \BC{} data only.
Using data below 2\,GeV/nucleon one may expect even tighter constraints to these parameters. 
Nevertheless, the parameters are determined with accuracy
$\delta D_{0}\sim\,0.5 \cdot 10^{28}$\,cm$^{2}$s$^{-1}$, $\delta L\sim\,0.5$\,kpc, 
and $\delta v_{A}\,\sim 2$\,km/s. This \emph{would} represent a great progress in CR propagation.

%
To study the impact of nuclear uncertainties, the procedure is now repeated after accounting 
for the estimated \XS{} errors.
In the $\chi^{2}$ calculation of Eq\,\ref{Eq::ChiSquare}, the \AMS{} errors are now summed in quadrature 
to the nuclear uncertainties $\delta$(\BC)$_{n}$ and  $\delta$(\BeB)$_{n}$.
The results are shown in Fig.\,\ref{Fig::ccChiSquaresAMS02}, bottom panel.
The nuclear uncertainties have an appreciable impact on the constraints provided by the \BC{} ratio,
and a dramatic impact in breaking the $D_{0}$ -- $L$ degeneracy.
In summary, this degeneracy remain unresolved when the nuclear uncertainties are taken into account. 
To lift the $D_{0}/L$ degeneracy, the information to be extracted in the \BeB{} ratio 
contained in the $^{10}$\Be$\rightarrow$$^{10}$\B{} decay, which produces only tiny variations in the \BeB{} ratio.
This information is blurred by the large uncertainties on the $^{10}$\Be{} production 
as well as by the uncertainties on the more abundant $^{7,9}$\Be{} and $^{11}$\B{} components.
Thus, a direct measurement of $^{10}$\Be{} at $\sim$\,1--10\,GeV/n would probably bring tighter constraints. 
To test this idea, the procedure was repeated after accounting the sole uncertainties in the $^{10}$\Be{} production.
In this case, the precision of the recontructed parameters is found to be
$\delta D_{0} \sim 1.5 \cdot 10^{28}\,$cm$^{2}$s$^{-1}$ and $\delta L \sim 1.5\;$kpc,
which still represents large uncertainties in comparisons to the \AMS{} potential.
However, given the unavoidable nuclear uncertainties of secondary production models, 
a direct measurement of $^{10}$\Be{} flux seems to bring much cleaner information than 
a precise \BeB{} measurement.
Besides the impact of the \XS{} uncertainties, it is also instructive to study the 
effect of systematic biases in single \PF{} reactions. 
This study and other elaborations connected with this work will be presented in a forthcoming work \cite{Tomassetti2015xsec}

\section{Conclusions and Discussions} 
%
These estimates show a promising potential for the \AMS{} experiment. 
\AMS{} is able to pose tight constraints on the key transport parameters,
thus we can expect a significant progress in CR propagation physics. 
Given its high level of precision the \emph{nuclear uncertainties} implicit in the models
are found to be a major limitation for the interpretation of the CR data.
After accounting for these uncertainties, the $D_{0}/L$ degeneracy remains poorly resolved
and the \BeB{} ratio appears to bring little information for the parameter extraction. 
Isotopically resolved $^{10}$\Be/$^{9}$\Be{} data would probably be preferable, 
though the $^{10}$\Be{} production rate is also affected by large uncertainties.
On the other hand, precise \BeB{} data at $E\gtrsim$\,10\,GeV/nucleon may represent a 
powerful tool for testing the nuclear physics inputs of the propagation models,
or to detect possible biases that may cause a parameter mis-determination.
It worth stressing that this problem has a direct impact for the dark matter search \citep{Gondolo2014}. 
In the \AMS{} era, the uncertainties of nuclear data have become a major limiting factor for further progress in CR propagation.
In this light, there is an urgent need for a dedicated experimental program of \XS{} measurements and modeling.

\end{document}